\documentclass[a4paper,11pt]{article}
\pdfoutput=1 
\usepackage{jheppub} 
\usepackage[T1]{fontenc} 
\usepackage{amsmath,amssymb,epsfig,ulem,color,slashed,wasysym}
\allowdisplaybreaks 

\def \beq{\begin{equation}}
\def \eeq{\end{equation}}
\def \bea{\begin{eqnarray}}
\def \eea{\end{eqnarray}}

\def\bm#1{\mbox{\boldmath$#1$\unboldmath}} 

\title{Indirect probes of the  trilinear Higgs coupling: $\bm{gg \to h}$ and $\bm{h \to \gamma \gamma}$}

\author[1]{Martin Gorbahn}
\author[2,3]{and Ulrich Haisch}

\affiliation[1]{Department of Mathematical Sciences, University of Liverpool, 
\\ L69 7ZL Liverpool, United Kingdom}

\affiliation[2]{Rudolf Peierls Centre for Theoretical Physics,
    University of Oxford, \\ OX1 3NP Oxford, United Kingdom}
    
\affiliation[3]{CERN, Theory Division, \\ CH-1211 Geneva 23, Switzerland}

\emailAdd{Martin.Gorbahn@liverpool.ac.uk}

\emailAdd{ulrich.haisch@physics.ox.ac.uk}

\abstract{In the framework of the Standard Model effective field theory, we examine the indirect constraints on the trilinear Higgs coupling $\lambda$ that arise from Higgs production in gluon-gluon-fusion and diphoton Higgs  decays. We calculate 2-loop contributions to the~$gg \to h$ and~$h \to \gamma \gamma$  amplitudes that are affected by modifications of the trilinear Higgs-boson vertex. This calculation involves both the computation of anomalous dimensions and finite matching corrections. Based on our new results, we analyse the sensitivity of present and future measurements of the $hgg$ and $h \gamma \gamma$ couplings to shifts in $\lambda$. Under the assumption that $O_6 = - \lambda \left (H^\dagger H \right )^3$ is the only  dimension-6 operator  that alters the trilinear Higgs  interactions, we find that at present the considered loop-level probes provide stronger constraints than $pp \to 2h$. At future high-energy colliders indirect ${\cal O} (5)$ determinations of the trilinear Higgs coupling may be possible, making precision measurements of $gg \to h$ and~$h \to \gamma \gamma$ a useful addition to direct extractions of $\lambda$ through double-Higgs production.}

\preprint{LTH 1092, CERN-TH-2016-161}

\begin{document} 

\maketitle

\flushbottom

\section{Introduction}
\label{sec:introduction}

The discovery of a boson and measurements of its production and decay rates at the Large Hadron Collider (LHC) give convincing evidence for the Higgs mechanism in which  a linearly realised $SU(2)_L \times U(1)_Y$ symmetry is spontaneously broken to $U(1)_{\rm EM}$ by the  vacuum expectation value (VEV) of the Higgs field. The new state seems to behave like a CP-even scalar and has couplings to the gauge bosons and fermions that agree with those predicted by the Standard Model (SM) at the level of 20\% to 100\% \cite{Aad:2015gba,Khachatryan:2014jba}. The strength of the Higgs couplings to the other SM fields will be further scrutinised at  forthcoming LHC runs  and~(possibly) other future high-energy colliders such as an International Linear Collider or a  Future Circular Collider~(FCC). 

The mechanism of electroweak symmetry breaking (EWSB) is in contrast to the Higgs gauge boson and fermion couplings essentially unexplored. Within the SM, the mass and the self-interactions  of the   physical Higgs field $h$ are parametrised by  the potential 
\beq \label{eq:VSM}
V_{\rm SM} = \frac{m_h^2}{2} \hspace{0.5mm} h^2 + \lambda v \hspace{0.25mm} h^3 + \frac{\kappa}{4} \hspace{0.5mm} h^4 \,, 
\eeq
where  $v \simeq 246 \, {\rm GeV}$ denotes  the Higgs VEV and 
\beq \label{eq:l34}
 \lambda = \kappa = \frac{m_h^2}{2v^2} \,.
\eeq
The LHC measurement  of the Higgs-boson mass leading to $m_h \simeq 125 \, {\rm GeV}$, probes the first term in (\ref{eq:VSM}), but the $h^3$ and $h^4$ couplings, and in particular the SM relation (\ref{eq:l34}) have not been tested. Determinations of the Higgs self-couplings are therefore an essential task that might provide indirect access to beyond the SM (BSM) dynamics, or if no significant deviations from (\ref{eq:l34}) are found, will further add to the impressive track record of the SM. 

High-energy collider allow  to probe the coefficients $\lambda$ and $\kappa$ in (\ref{eq:VSM}) through double-Higgs and triple-Higgs production, respectively. At  $14 \, {\rm TeV}$ centre-of-mass energy, the cross section for $pp \to 3h$ production is of ${\cal O} (0.1 \,{\rm fb})$, which even at the high-luminosity LHC~(HL-LHC) renders any meaningful extraction of $\kappa$  impossible (see for instance \cite{Plehn:2005nk,Binoth:2006ym,Maltoni:2014eza}). The prospect to observe double-Higgs production at the HL-LHC is significantly better since at~$14 \, {\rm TeV}$ the $pp \to 2h$ production cross section amounts to~${\cal O} (35 \,{\rm fb})$~\cite{Glover:1987nx,Plehn:1996wb,Dawson:1998py,Djouadi:1999rca,deFlorian:2013jea,Grigo:2013rya}. LHC detections of double-Higgs production still remain challenging~(cf.~\cite{Baur:2002qd,Baur:2003gp,Dolan:2012rv,Baglio:2012np,Barr:2013tda,Dolan:2013rja,Papaefstathiou:2012qe,Goertz:2013kp,Maierhofer:2013sha,deLima:2014dta,Englert:2014uqa,Liu:2014rva,Goertz:2014qta,ATL-PHYS-PUB-2014-019,Azatov:2015oxa,Dall'Osso:2015aia,ATL-PHYS-PUB-2015-046}) and in consequence even with the full HL-LHC data set of~$3 \, {\rm ab}^{-1}$ only  an ${\cal O} (1)$ determination of the trilinear Higgs coupling seems feasible under optimistic assumptions. 

This raises the question if it is possible to constrain~$\lambda$ by other~(possibly complementary) means. As pointed out in \cite{McCullough:2013rea}, one way to achieve sensitivity to the $h^3$ coupling is provided through precision studies of the process~$e^+ e^- \to hZ$, which receives 1-loop corrections proportional to~$\lambda$ from Feynman diagrams with virtual Higgses and/or a $Z$ boson (see more recently also~\cite{Shen:2015pha}). While loop-level probes of the~$h^3$ vertex can clearly not replace extractions of~$\lambda$ based on collider measurements of double-Higgs production, they might be crucial in resolving degeneracies in parameter space. This is evident by recalling that $gg \to 2h$ production is itself a quantum process that depends not only on the strength of the $h^3$ interaction, but also on the top-quark Yukawa coupling as well as all the masses and all the Higgs couplings of the possible BSM particles circulating in the loop graphs. 

The main purpose of our work is to illustrate other ways to indirectly probe the coefficient~$\lambda$ entering (\ref{eq:VSM}). To keep our discussion as model-independent as possible, we will use an effective field theory (EFT) in which the SM Lagrangian is the leading term and BSM effects are encoded in dimension-6 operators constructed solely out of SM fields. In this framework, we calculate  2-loop contributions to the $gg \to h$ and $h \to \gamma \gamma$ amplitudes that are affected by modifications of the $h^3$ vertex. This calculation involves both the computation of the relevant anomalous dimensions of the operators as well as the computation of the finite matching corrections at the weak scale. Based on our results, we analyse the sensitivity of present and future measurements of the $hgg$ and the $h\gamma\gamma$ couplings to shifts in the trilinear Higgs interactions. We find  that our new loop-level probes provide interesting and meaningful  model-independent constraints on $\lambda$, in the simplified case that the operator~$O_6 = - \lambda \left (H^\dagger H \right )^3$ furnishes the sole modification of the  $h^3$ vertex.  Applying our findings to ultraviolet (UV) complete realisations of BSM physics is left for further study. 

This article is structured as follows. In Section~\ref{sec:preliminaries} we introduce the  effective interactions relevant for our paper. The results of our 2-loop mixing and matching calculations are presented in Section~\ref{sec:anomalousdimensions} and \ref{sec:results}, respectively. Our numerical analyses are performed in Sections~\ref{sec:numerics1} and~\ref{sec:numerics2}. We conclude in Section~\ref{sec:conclusions}. Some technical details of our computations are described in Appendix~\ref{app:EOM}, \ref{app:Ophi}, \ref{app:matching} and \ref{app:goldstones}.

\section{Preliminaries}
\label{sec:preliminaries}

New physics can be described in a model-independent way by augmenting the SM Lagrangian ${\cal L}_{\rm SM}$ by  $SU(3)_C \times SU(2)_L \times U(1)_Y$  gauge-invariant higher-dimensional operators. In our  work, we consider the effective Lagrangian 
\beq \label{eq:Ldim6}
{\cal L}_{\rm EFT}= \sum_{k} \frac{\bar c_k}{v^2}  \, O_k \,,
\eeq
built out of the following dimension-6 operators
\beq \label{eq:SILHoperator}
\begin{split}
O_6 & = -\lambda \, \big ( H^\dagger H \big )^3 \,, \\[1mm]
O_H & = \frac{1}{2}  \, \partial_\mu \big ( H^\dagger H \big ) \, \partial^\mu \big ( H^\dagger H \big ) \,, \\[1mm]
O_T & = \frac{1}{2}  \,  \big ( H^\dagger \! \stackrel{\leftrightarrow}{D}_\mu \! H  \big )  \big ( H^\dagger   {\stackrel{\leftrightarrow}{D}}^{\, \mu}  H \big ) \,,  \\[1mm]
O_W & =  \frac{4 i}{g} \, \big ( H^\dagger \tau^i \! \stackrel{\leftrightarrow}{D}_\mu \! H  \big ) D_\nu W^{i, \mu \nu} \,, \\[1mm]
O_B & =  \frac{2 i g^\prime }{g^2} \, \big ( H^\dagger \! \stackrel{\leftrightarrow}{D}_\mu \! H  \big ) D_\nu B^{\mu \nu} \,, \\[1mm]
O_{HW} & =  \frac{8 i}{g} \, \big ( D_\mu H^\dagger \tau^i D_\nu H \big )  W^{i, \mu \nu} \,, \\[1mm]
O_{HB} & =  \frac{4 i  g^\prime}{g^2} \, \big ( D_\mu H^\dagger D_\nu H \big )  B^{\mu \nu} \,, \\[1mm]
O_{GG}  & =  \frac{2 g_s}{g^2} \, H^\dagger H \, G^a_{\mu \nu} G^{a, \mu \nu} \,, \\[1mm]
O_{BB}  & =  \frac{2 g^\prime}{g^2} \, H^\dagger H \, B_{\mu \nu} B^{\mu \nu} \,, \\[1mm]
O_u & = - Y_u \, H^\dagger H \, \bar Q_L  u_R \hspace{0.25mm} \tilde H\,, \\[2mm]
O_d & = - Y_d \, H^\dagger H \, \bar Q_L d_R  \hspace{0.25mm} H \,, \\[2mm] 
O_\ell & = - Y_\ell \, H^\dagger H \, \bar L_L   \ell_R  \hspace{0.25mm} H \,. 
\end{split}
\eeq
Here $\lambda$ denotes the SM Higgs self-coupling introduced in (\ref{eq:l34}), $H$ is the SM Higgs doublet and we have used the  shorthand notation $\tilde H^i = \epsilon_{ij} \left ( H^{j}  \right)^\ast$ with $\epsilon_{ij}$ totally antisymmetric and $\epsilon_{12}=1$. The covariant derivative operator $\stackrel{\leftrightarrow}{D}_\mu$ is defined as $H^\dagger \! \stackrel{\leftrightarrow}{D}_\mu \! H = H^\dagger D_\mu H - \big ( D_\mu H^\dagger \big) H$ and~$\tau^i = \sigma^i/2$ with $\sigma^i$ the usual Pauli matrices. The coupling constants of the gauge groups~$SU(3)_C$, $SU(2)_L$ and $U(1)_Y$ are denoted by $g_s$, $g$ and $g^\prime$, while $G^a_{\mu \nu}$, $W_{\mu \nu}^i$, and~$B_{\mu \nu}$ are the corresponding field strength tensors. The Yukawa couplings $Y_u$, $Y_d$ and $Y_\ell$ are matrices in flavour space and a sum over flavours indices is implicit in (\ref{eq:SILHoperator}). Finally, $Q_L, L_L$ denote left-handed quark and lepton doublets, while $u_R, d_R, \ell_R$ are right-handed fermion singlets. 

After EWSB, the $SU(3)_C \times SU(2)_L \times U(1)_Y$ gauge-invariant operators introduced in~(\ref{eq:SILHoperator}) modify the couplings of the Higgs boson to itself, to vector bosons and fermions. We write the couplings that result from ${\cal L}_{\rm SM} + {\cal L}_{\rm EFT}$ and that are relevant for our article as follows 
\beq \label{eq:Ltriple}
{\cal L}  \supset  - \lambda c_3  \hspace{0.25mm} v  \hspace{0.25mm} h^3  + c_g \, \frac{h}{v} \, G_{\mu \nu}^a G^{a, \mu \nu} 
+ c_\gamma \, \frac{h}{v} \, F_{\mu \nu} F^{\mu \nu} \,,
\eeq
where $F_{\mu \nu} = \partial_\mu A_\nu - \partial_\mu A_\nu$ with $A_\mu$  the photon field. Upon canonical normalisation of the Higgs kinetic term the tree-level coefficient $c_3$ takes the form
\beq \label{eq:treecoefficients}
c_3  = 1 + \bar c_6 - \frac{3 \bar c_H}{2} \,.
\eeq
The coefficients $c_g$ and $c_\gamma$ arise first at the 1-loop level and we will give the relevant expressions below. Note finally that the Wilson coefficients $c_k$ as well as the trilinear Higgs coupling~$\lambda$ appearing in~(\ref{eq:Ltriple}) are all understood to be evaluated at the weak scale, which we will denote by $\mu_w$ hereafter. 

\section{Anomalous dimensions}
\label{sec:anomalousdimensions}

The primary goal of this article is to determine the dominant corrections to the Higgs couplings to gluon and photon pairs (\ref{eq:Ltriple}) that result from the effective operator $O_6$. In fact,  such contributions  can arise in two ways. First, via renormalisation group (RG) evolution 
\beq \label{eq:RGE}
\frac{d \bar c_k}{d \ln \mu} = \gamma_{k6} \, \bar c_{6} \,,
\eeq
of the Wilson coefficients $\bar c_k$ from the new-physics scale $\Lambda$ down to $\mu_w$, if the operator  $O_6$ mixes into $O_k$. Second, from matching Green's functions  obtained in the theory described by ${\cal L}_{\rm SM} + {\cal L}_{\rm EFT}$ to those resulting from ${\cal L}$. In the following, we discuss the corrections associated to the RG evolution, turning our attention to the matching corrections in the next section. 

Since a non-zero initial condition $\bar c_6$ at $\Lambda$ does not affect the other weak-scale Wilson coefficients $\bar c_k$ at the 1-loop level \cite{Elias-Miro:2013gya,Jenkins:2013zja,Jenkins:2013wua,Alonso:2013hga}, the leading logarithmic corrections to (\ref{eq:Ltriple}) proportional to $\bar c_6$ have to arise from 2-loop diagrams. In order to determine the mixing of $O_6$ into the set of operators introduced in (\ref{eq:SILHoperator}) we have calculated the~2-loop matrix elements $HH\to HH$, $HH \to BB$, $HH \to WW$, $HH \to BW$ and $HHH \to f \bar f$ involving a single insertion of~$O_6$.  The pole  parts of the graphs have been evaluated using the method described for instance in~\cite{Chetyrkin:1997fm,Gambino:2003zm,Gorbahn:2004my}. Specifically, we have performed the calculation off-shell in an arbitrary~$R_\xi$ gauge which allows us to explicitly check the $\xi$-independence of the mixing among physical operators. To distinguish between infrared~(IR) and UV divergences,  a common mass $M$ for all fields is introduced, expanding the loop integrals in inverse powers of $M$. This makes the calculation of the  2-loop UV  divergences straightforward, because after Taylor expansion in the external momenta, $M$ becomes the only relevant internal scale and 2-loop tadpole integrals with a single non-zero mass are known~\cite{Avdeev:1994db}.  Further  technical details on our off-shell calculation are given in Appendix~\ref{app:EOM}. 

We find that the only non-vanishing anomalous dimensions $\gamma_{k6}$ that encode the 2-loop off-diagonal mixing of $O_6$ into the operators of (\ref{eq:SILHoperator}) are
\beq \label{eq:twoloopADM} 
\gamma_{H6}  = \frac{1}{16 \pi^4}\; 12 \hspace{0.25mm} \lambda^2  \,, \qquad 
\gamma_{f 6}  = -\frac{1}{16 \pi^4}  \left ( \lambda^2 +  3 \hspace{0.5mm} Y_f Y_f^\dagger \right ) \,, 
\eeq
where $f = u, d, \ell$. These results imply that  the weak-scale Wilson coefficients of  $O_H$ and~$O_f$ alone receive logarithmically-enhanced contributions $\ln \left (\Lambda^2/\mu_w^2 \right )$ proportional to $\bar c_6$ at the 2-loop level. In this context it is also important to realise  that the higher-dimensional interactions introduced in (\ref{eq:SILHoperator}) provide just  a subset   of the dimension-6 operators of the full SM effective Lagrangian~(cf.~\cite{Buchmuller:1985jz,Grzadkowski:2010es}). In particular, operators that are composed out of three field strength tensors such as $O_{3W} = 4 g^2 \, \epsilon_{ijk} W_{\mu \nu}^i W^{j, \nu}_{\rho} W^{k, \rho \mu}$ with $\epsilon_{ijk}$ the Levi-Civita tensor are not included in ${\cal L}_{\rm EFT}$.  Since $O_6$ involves three powers of $H^\dagger H$ it however cannot give rise to amplitudes like $W \to WW$ at two loops, because one has to contract all $H$ fields to obtain a non-zero matrix element. Since this  is first possible at the 3-loop level, all 2-loop  anomalous dimensions describing the mixing of $O_6$ into dimension-6 operators containing only field strength tensors vanish identically. Beyond that order such mixings are likely to be present, but a computation of these logarithmic 3-loop corrections is beyond the scope of this work. 

\section{Matching corrections for $\bm{gg \to h}$ and $\bm{h \to \gamma \gamma}$}
\label{sec:results}

As already mentioned, a second type of contributions to the coefficients $c_g$ and $c_\gamma$ entering~(\ref{eq:Ltriple}) stems from matching Green's functions obtained in the theory described by~${\cal L}_{\rm SM} + {\cal L}_{\rm EFT}$ to those originating from ${\cal L}$. 

We first discuss the corrections arising in the case of the $gg \to h$ amplitude. Expanding the corresponding Wilson coefficient $c_g$  as follows 
\beq \label{eq:cgexpansion}
c_g = \frac{\alpha_s}{\pi} \left ( c_g^{(0)} +  \frac{\lambda}{(4 \pi)^2} \, c_g^{(1)} \right ) \,, 
\eeq
one obtains at the 1-loop level 
\beq \label{eq:cg0}
c_g^{(0)} = \sum_q  A_q  \simeq 0.081 + 0.007 \hspace{0.5mm} i  \,, 
\eeq
where the sum runs over all quarks and 
\beq \label{eq:Af}
A_f = \frac{\tau_f}{8} \left [ 1 + ( 1- \tau_f) \arctan^2 \frac{1}{\sqrt{\tau_f - 1}} \right ] \,,
\eeq
with $\tau_f = 4 m_f^2/m_h^2$.  The numerical value given in (\ref{eq:cg0}) corresponds to $m_t \simeq 163.3 \, {\rm GeV}$, $m_b \simeq 4.2 \, {\rm GeV}$, $m_c \simeq 1.3  \, {\rm GeV}$ and $m_h \simeq 125 \, {\rm GeV}$. Since the  on-shell 1-loop form factor  $A_f$  approaches $1/12$ for $\tau_f \to \infty$ and vanishes proportional to $\tau_f$ in the limit $\tau_f \to 0$, it is an excellent approximation to include only the top quark in the sum appearing in~(\ref{eq:cg0}) and to take the infinite quark-mass limit. In such a case, one arrives at the  classic Shifman-Vainshtein-Zakharov result $c_g^{(0)} =1/12 \simeq 0.083$ derived first in~\cite{Shifman:1978zn}.

\begin{figure}[!t]
\begin{center}
\includegraphics[width=0.45\textwidth]{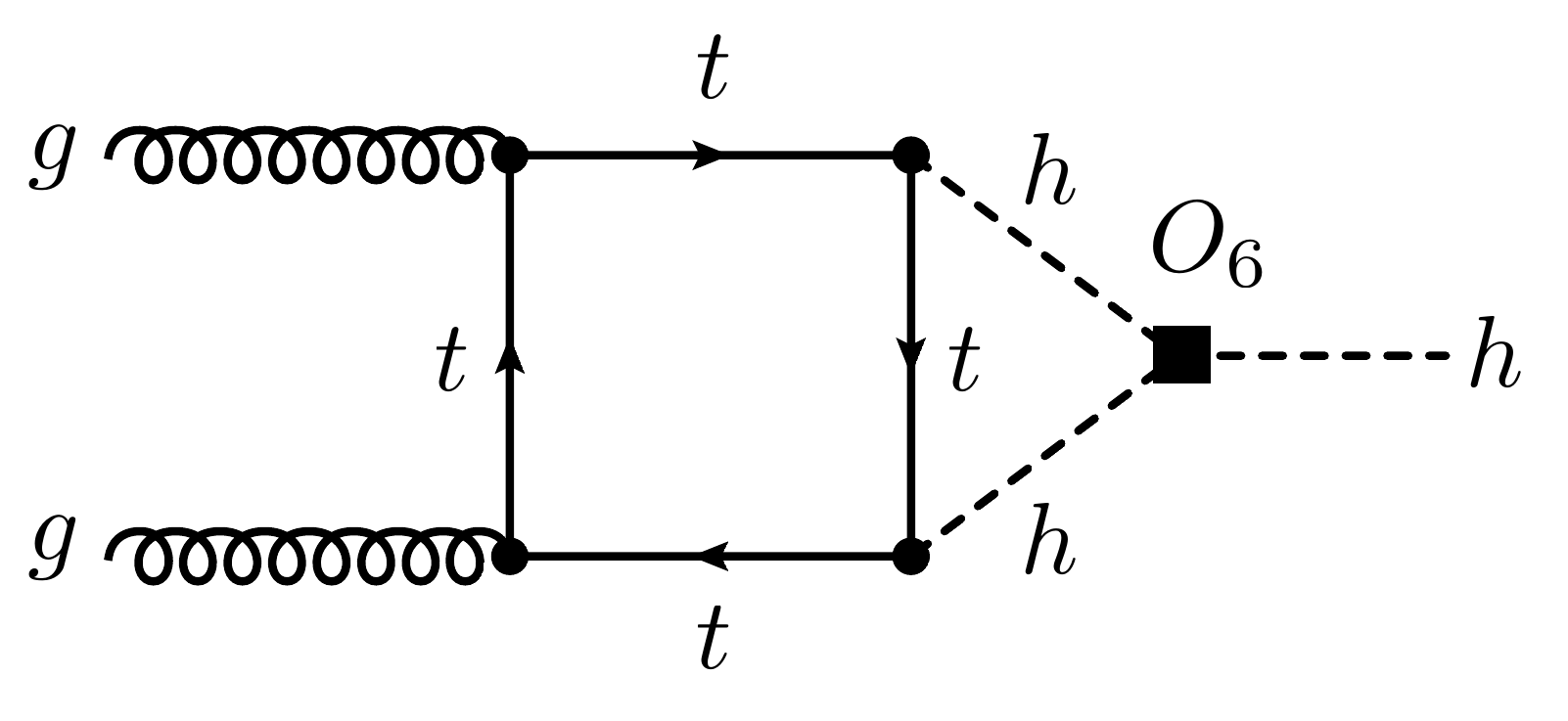}  
\vspace{2mm}
\caption{\label{fig:gghdiagrams} Example of a 2-loop diagram with an insertion of the effective operator $O_6$ that contributes to the $gg \to h$ amplitude at ${\cal O}(\lambda)$.}
\end{center}
\end{figure}

The ${\cal O} (\lambda)$ correction to the coefficient $c_g$ arises from both 2-loop Feynman diagrams and 1-loop counterterm graphs involving a Higgs wave function renormalisation. To find the former type of contribution, we apply EFT techniques (see for instance  \cite{Anastasiou:2008tj} for a non-trivial application to  Higgs production) and employ a hard-mass expansion procedure $\tau_t \to \infty$ to the full 2-loop diagrams involving a top-quark loop and a~$h^3$ vertex that arises from the insertion of $O_6$. A prototype  graph of such a  contribution is shown in Figure~\ref{fig:gghdiagrams}. After setting $m_h = 0$ and Taylor expanding in the external momenta, this technique reduces the calculation to the evaluation of 2-loop vacuum bubbles with a single mass scale, which can all be expressed in terms of Gamma functions~(cf.~\cite{Avdeev:1994db}). 

The correction proportional to the ${\cal O} (\lambda)$ contribution to the Higgs wave function renormalisation constant 
\beq \label{eq:Zhexpand}
Z_h = 1 + \frac{\lambda}{(4 \pi)^2} \, Z_h^{(1)} \,,
\eeq
is instead found  from the 1-loop Higgs-boson selfenergy with one and two insertions of~$O_6$. By a straightforward calculation, we obtain the analytic result
\beq \label{eq:Zh1}
Z_h^{(1)} = \left ( 9 - 2 \sqrt{3} \pi \right ) \bar c_6 \left (\bar c_6 + 2 \right ) \,.
\eeq 

Combining both contributions, we arrive at
\beq \label{eq:cg1}
c_g^{(1)} = -\frac{1}{12} \left ( \frac{1}{4} + 3 \ln \frac{\mu_w^2}{m_t^2} \right ) \bar c_6 + \frac{Z_h^{(1)}}{2} \, c_g^{(0)}  \,, 
\eeq 
with $c_g^{(0)}$ given in (\ref{eq:cg0}).  As a powerful cross-check of our calculation, we have extracted the~${\cal O} (\lambda)$  correction to the coefficient $c_g$ arising from 2-loop diagrams  by matching in addition the $gg \to 2h$ and $gg \to 3h$ Green's functions, obtaining in all three cases the exact same result. Details on the renormalisation of the bare 2-loop $gg \to h$ amplitude can be found in Appendix~\ref{app:matching}. Given the good convergence of the infinite quark-mass expansion in the case of $c_g^{(0)}$, we believe that our analytic expression (\ref{eq:cg1}) should approximate the full~${\cal O} (\lambda)$ correction to the on-shell 2-loop form factor quite well. To make this statement more precise would require an explicit calculation of the relevant $gg \to h$ amplitudes that does not rely on the heavy-quark expansion for what concerns the 2-loop contributions.  Such a computation is however beyond the scope of our article. 

In the case of the $h \to \gamma\gamma$ transition, we write  
\beq \label{eq:cAexpansion}
c_\gamma = \frac{\alpha}{\pi} \left ( c_\gamma^{(0)} +  \frac{ \lambda}{(4 \pi)^2} \, c_\gamma^{(1)} \right ) \,,
\eeq
where the 1-loop contribution is given by 
\beq \label{eq:cA0}
c_\gamma^{(0)} = A_W + \sum_f   2 N_C^f \hspace{0.25mm} Q_f^2 \hspace{0.25mm} A_f  \simeq -0.82 -0.01 \hspace{0.5mm} i  \,.
\eeq
Here $N_C^q =3$ and $N_C^\ell =1$ are colour factors, the sum runs over all electrically charged fermions carrying charge~$Q_u=2/3$, $Q_d=-1/3$ and $Q_\ell =-1$, $A_f$ has been introduced in~(\ref{eq:Af}) and 
\beq \label{eq:AW}
A_W = -\frac{1}{8} \left [ 2+ 3\tau_W + 3 \tau_W (2-\tau_W) \arctan^2 \frac{1}{\sqrt{\tau_W-1}} \right ] \,,
\eeq
with $\tau_W = 4 m_W^2/m_h^2$.  In order to obtain the numerical result in (\ref{eq:cA0}), we have employed $m_W \simeq 80.4 \, {\rm GeV}$ and $m_\tau \simeq 1.777 \, {\rm GeV}$.  Numerically, one has furthermore $A_W \simeq -1.04$, while in the limit $\tau_W \to \infty$ ($\tau_W \to 0$) the on-shell 1-loop form factor $A_W$ tends to the constant value $-7/8$~($-1/4$). In the infinite mass limit $\tau_{t,W} \to \infty$, one therefore finds that $c_\gamma^{(0)} = -47/72 \simeq -0.65$. Notice that compared to the case of $A_f$  the heavy-mass expansion works less well for~$A_W$, but still captures around $85\%$ of the exact 1-loop result. We thus believe that the hard-mass expansion is also a sufficiently accurate approximation in the case of the 2-loop corrections to $c_\gamma$ involving $W^\pm$ ($\phi^\pm$) exchanges. 

\begin{figure}[!t]
\begin{center}
\includegraphics[height=0.215 \textwidth]{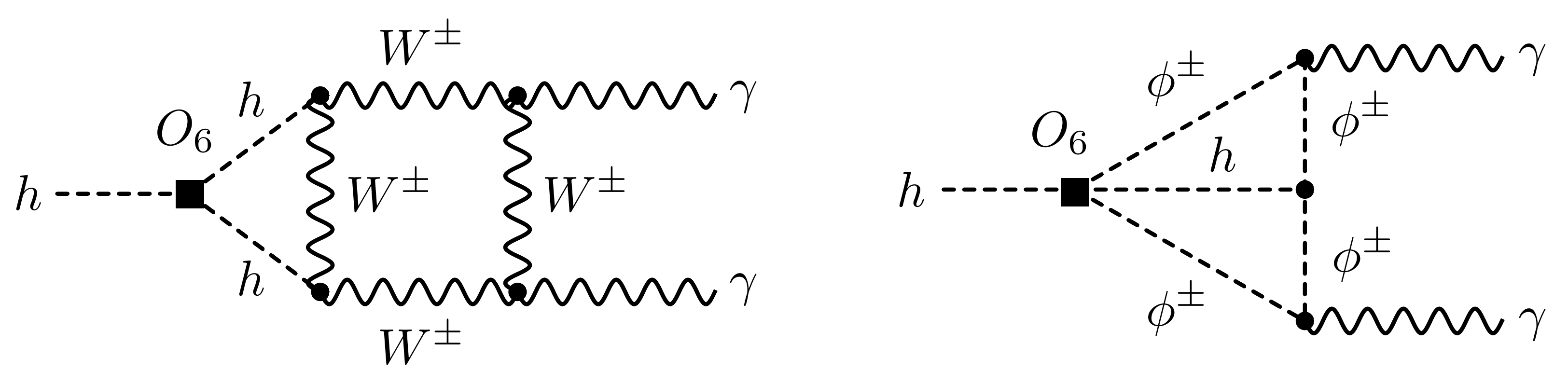} 
\vspace{2mm}
\caption{\label{fig:haadiagrams} Left: Example of a 2-loop diagram involving a $h^3$ vertex that in the limit $m_W \gg m_h$ gives rise to $h \to \gamma \gamma$  at ${\cal O} (\lambda)$. Right: A possible 2-loop graph with a $h \phi^+ \phi^-$ vertex. For $m_h = 0$, diagrams of this type do however not contribute  to $h \to \gamma \gamma$ at ${\cal O} (\lambda)$. For additional explanations see text. 
}
\end{center}
\end{figure}

Since after EWSB the operator $O_6$ modifies both the trilinear Higgs coupling as well as the coupling between two Higgses and two charged would-be Goldstone bosons $\big($see (\ref{eq:Ophi})$\big)$, one naively has to consider 2-loop diagrams that contain both a $h^3$ and a~$h^2 \phi^+ \phi^-$ vertex. A possible graph of each type is depicted in  Figure~\ref{fig:haadiagrams}. To maintain gauge invariance at the level of off-shell Green's functions, we use the 't Hooft-Feynman version of the background field gauge for the external photon fields (see~e.g.~\cite{Denner:1994xt}) when calculating these diagrams. In this gauge there is no $\gamma W^\pm \phi^\mp$ vertex and as a result all 2-loop graphs involving a $h^2 \phi^+ \phi^-$ interaction necessarily also contain a~$h \phi^+ \phi^-$ vertex. Since the Feynman rule of the~$h \phi^+ \phi^-$ coupling is proportional to $m_h^2/m_W$, it then follows that diagrams with a~$h^2 \phi^+ \phi^-$ vertex do not contribute to  $c_\gamma^{(1)}$ in the  limit $\tau_W \to \infty$. We have verified this feature by an explicit calculation of the corresponding 2-loop graphs. 

Combining the 2-loop gauge boson and  top-quark contributions and including the correction due to the wave function renormalisation of the Higgs boson, we find for $\tau_{t,W} \to \infty$ the following analytic expression
\beq \label{eq:cA1}
c_\gamma^{(1)} =  -\frac{7}{8} \left ( \frac{9}{7} - \frac{22}{7} \ln \frac{\mu_w^2}{m_W^2} \right )  \bar c_6    -\frac{2}{9} \left ( \frac{1}{4} + 3 \ln \frac{\mu_w^2}{m_t^2} \right ) \bar c_6 + \frac{Z_h^{(1)}}{2} \, c_\gamma^{(0)}  \,,
\eeq 
with  $Z_h^{(1)}$ and $c_\gamma^{(0)}$ given in (\ref{eq:Zh1}) and (\ref{eq:cA0}), respectively. The necessary ingredients to obtain the above result are presented in Appendix~\ref{app:matching}, while the renormalisation of the would-be Goldstone boson sector is discussed in Appendix~\ref{app:goldstones}.

\section{Constraints from double-Higgs production}
\label{sec:numerics1}

In the next section will derive existing and possible future limits on the modifications of the~$h^3$ coupling that arise from $gg \to h$ and $h \to \gamma \gamma$. All the numbers that we will present below should be compared to the bounds  on the trilinear Higgs coupling that one can obtain by studying double-Higgs production at the LHC. 

In fact, ATLAS  has combined the full $8 \, {\rm TeV}$ data set corresponding to $20.3 \, {\rm fb}^{-1}$ of integrated luminosity to search for $2h \to 2b 2\tau, 2 \gamma 2 W, 2 \gamma 2 b$ and $4 b$ \cite{Aad:2014yja, Aad:2015uka, Aad:2015xja}. While no evidence for double-Higgs production is observed, a 95\% confidence level~(CL) upper limit of~$0.69 \, {\rm pb}$ is set on the production cross section, which is about $70$ times above the SM expectation of $(9.9 \pm 1.3) \, {\rm fb}$ \cite{deFlorian:2013jea}. To derive a bound on $\bar c_6$, we use the {\tt MadGraph5\_aMC@NLO}~\cite{Alwall:2014hca} implementation of cross section computations for loop-induced processes \cite{Hirschi:2015iia} as well as {\tt HPAIR}~\cite{Grober:2015cwa,hpair}. For $pp$ collisions at $8 \, {\rm TeV}$, we obtain 
\beq
\sigma (p p \to 2 h) = \left ( 9.9 \pm 1.3 \right ) \left ( 1 - 0.87 \hspace{0.25mm} \bar c_6 + 0.33 \hspace{0.25mm} \bar c_6^{\hspace{0.5mm} 2} \right ) {\rm fb} \,,
\eeq
if only the Wilson coefficient $\bar c_6$ is allowed to be non-zero, but the remaining coefficients~$\bar c_k$ of the operators entering (\ref{eq:SILHoperator}) are assumed to vanish. Taking into account theoretical uncertainties, we find from this formula that the  ATLAS limit on the $pp \to 2h$ production cross section translates into the following~95\%~CL bound
\beq \label{eq:cphiATLAS}
\bar c_6 \in [-18.2, 15.6] \,.
\eeq
This limit implies that the combination $\lambda \bar c_3$ introduced in (\ref{eq:Ltriple}) can at present still deviate from the SM trilinear Higgs coupling $\lambda$ by a factor of about $17$. This finding agrees with the conclusion drawn in \cite{Dall'Osso:2015aia}. 

At the HL-LHC with $3 \, {\rm ab}^{-1}$ of integrated luminosity the constraints on the trilinear Higgs coupling are expected to  improve  considerably (cf.~\cite{Baur:2002qd,Baur:2003gp,Dolan:2012rv,Baglio:2012np,Barr:2013tda,Dolan:2013rja,Papaefstathiou:2012qe,Goertz:2013kp,Maierhofer:2013sha,deLima:2014dta,Englert:2014uqa,Liu:2014rva,Goertz:2014qta,ATL-PHYS-PUB-2014-019,Azatov:2015oxa,Dall'Osso:2015aia,ATL-PHYS-PUB-2015-046}). For example, the analysis~\cite{Goertz:2014qta} implies that the ultimate sensitivity that the LHC can reach in the $2 h \to 2b 2\tau$ channel, will allow to set a 95\% CL bound of \cite{flogo}
\beq \label{eq:cphiHLLHCdirect}
\bar c_6 \in [-0.9,1.6] \cup [4.5,6.9]  \,,
\eeq
on the coefficient of the dimension-6 operator $O_6$. Slightly more pessimistic results have been obtained in the preliminary ATLAS study~\cite{ATL-PHYS-PUB-2015-046}.  In order to allow for a better comparison with (\ref{eq:cphiATLAS}) and the bounds derived in the next section, the quoted limit again assumes that~$\bar c_6$ is the only numerically relevant Wilson coefficient at the weak scale. If this assumption is relaxed the limit (\ref{eq:cphiHLLHCdirect}) can worsen by a factor of a few \cite{Goertz:2014qta,Azatov:2015oxa}. Notice finally that (\ref{eq:cphiHLLHCdirect}) exhibits two solutions. The first one is located close to the SM point at $\bar c_6 = 0$, while the second solution at $\bar c_6 \simeq 5.7$ corresponds to the case where the $gg \to 2 h$ amplitude has an opposite sign with respect to the SM. Removing the non-SM solution seems challenging at the HL-LHC, but should be possible at a $100 \, {\rm TeV}$ FCC-pp~\cite{Azatov:2015oxa}. 

\section{Constraints from Higgs production and diphoton decay}
\label{sec:numerics2}

In the following, we study the present constraints  and the future sensitivities on the trilinear Higgs coupling that are provided by $gg \to h$ and $h \to \gamma \gamma$. In order to allow for an easy comparison with the results~(\ref{eq:cphiATLAS}) and (\ref{eq:cphiHLLHCdirect}), we will throughout assume that the modifications of the Wilson coefficient of the operator $O_6$ furnish the dominant contribution to the observable under consideration, and consequently neglect effects associated  to the remaining $\bar c_k$ in  (\ref{eq:Ldim6}). 

The ratio of the cross sections for Higgs-boson production in gluon-gluon fusion and  the modification of the signal strength for Higgs decays into two photons  can be written as
\beq \label{eq:mugg} 
\mu_{gg} = \frac{\sigma (gg \to h)}{\sigma_{\rm SM} (gg \to h)} = |\kappa_g|^2\,, \qquad 
\mu_{\gamma\gamma} = \frac{\Gamma ( h \to \gamma \gamma)}{\Gamma_{\rm SM} ( h \to \gamma \gamma)} = | \kappa_\gamma |^2 \,,
\eeq
respectively. From the definitions (\ref{eq:cgexpansion}) and (\ref{eq:cAexpansion}) it is then readily seen that ($i = g, \gamma$)
\beq \label{eq:kappagamma}
\kappa_i \simeq 1 + \frac{\lambda}{(4\pi)^2} \, \frac{{\rm Re} \,  c_i^{(1)}}{{\rm Re} \, c_i^{(0)}}\,,
\eeq
where we have neglected the small imaginary parts of  $c_i^{(0)}$ and $c_i^{(1)}$. 

In order to set limits on the Wilson coefficient $\bar c_6$, we use the latest results of a global fit to the Higgs production channels performed by ATLAS~\cite{Aad:2015gba} and CMS~\cite{Khachatryan:2014jba}, where the effective couplings $\kappa_g$, $\kappa_\gamma$ and $\kappa_{\gamma Z}$ are left to vary freely. All the remaining couplings are set to their SM values. ATLAS and CMS obtain $\kappa_g = 1.12\pm 0.12$, $\kappa_\gamma = 1.00 \pm 0.12$  and $\kappa_g = 0.89 \pm 0.10$, $\kappa_\gamma = 1.15 \pm 0.13$, respectively. Performing a naive weighted average, one finds
\beq \label{eq:kappaggpresent}
\kappa_g = 0.98 \pm 0.08 \,, \qquad  \kappa_\gamma = 1.07 \pm 0.09 \,.
\eeq
Employing now (\ref{eq:cg0}), (\ref{eq:cg1}), (\ref{eq:cA0}), (\ref{eq:cA1}), identifying $\mu_w = m_h \simeq 125 \, {\rm GeV}$ and treating the extractions of $\kappa_g$ and $\kappa_\gamma$ as uncorrelated, these limits translate into 
\beq \label{eq:cphipresent}
 \bar c_6 \in [-12.7, 9.9] \,,
\eeq
at 95\% CL. One observes that the present indirect constraint arising from a combination of the observed $gg \to h$ and $h \to \gamma \gamma$ signal strengths is  more restrictive than  the direct bound~(\ref{eq:cphiATLAS}) from $pp \to 2h$ production. We believe that this is an interesting finding, because it shows that  it is possible to constrain the $h^3$ couplings at a $pp$ collider by means other  than double-Higgs production. 

Since at the time the LHC has collected $3 \, {\rm ab}^{-1}$ of data, the effective $hgg$ and $h \gamma \gamma$ couplings will be known much more accurately than today as well, it is also interesting to study the prospects of the indirect probes provided by $gg \to h$ and $h \to \gamma \gamma$. The sensitivity study~\cite{NOTE2012-006}  finds for instance that compared to (\ref{eq:kappaggpresent}) the precision on $\kappa_g$ ($\kappa_\gamma$) might be improved by a  factor of 3 (4). Assuming that the central values of the future LHC measurements end up being spot on the SM, this means 
\beq \label{eq:kappasHLLHC}
\kappa_g = 1.00 \pm 0.03 \,, \qquad  \kappa_\gamma = 1.00 \pm 0.02 \,.
\eeq
The corresponding 95\% CL limit on the Wilson coefficient of $O_6$ reads
\beq \label{eq:cphiHLLHC}
 \bar c_6 \in [-8.0, 5.1] \,.
\eeq
With such a precision one will start to become sensitive to the flipped-sign solution in (\ref{eq:cphiHLLHCdirect}). 

At a  $e^+ e^-$ option of a FCC (FCC-ee), the bounds (\ref{eq:kappasHLLHC}) may even be further tightened~\cite{Ruan:2014xxa}, possibly leading to 
\beq \label{eq:kappasFCCee}
\kappa_g = 1.00 \pm 0.01 \,, \qquad  \kappa_\gamma = 1.00 \pm 0.015 \,.
\eeq
With this sensitivity, one could set the following 95\% CL limit
\beq \label{eq:cphiFCCee}
 \bar c_6 \in [-5.3, 3.8] \,.
\eeq
This bound improves on the LHC Run~I constraint (\ref{eq:cphipresent}) by a factor of around 2.5. Notice that  in this case the non-SM solution present in (\ref{eq:cphiHLLHCdirect}) would be fully removed by the combination of $gg \to h$ and $h \to \gamma \gamma$. 
 
\section{Conclusions}
\label{sec:conclusions}

In this article, we have proposed to constrain deviations in the trilinear Higgs coupling $\lambda$ by studying Higgs production in gluon-gluon-fusion and diphoton Higgs decays. To keep our discussion general, we have employed the SM EFT, in which new-physics effects are described by dimension-6 operators. In this framework, we have calculate~2-loop contributions to  the~$gg \to h$ and~$h \to \gamma \gamma$ amplitudes that are affected by insertions of the effective operator $O_6 = -\lambda \left (H^\dagger H \right)^3$. By an explicit calculation of the complete~2-loop anomalous dimensions involving a single insertion of $O_6$, we have  shown that the effective~$hgg$ and~$h\gamma\gamma$  couplings do not receive logarithmically-enhanced contributions proportional to $\bar c_6$. The leading  contributions to the $gg \to h$ and $h \to \gamma \gamma$ transitions involving  the Wilson coefficient of $O_6$ hence arise from  finite 2-loop matching corrections at the weak scale. We have calculated these corrections by employing a heavy-mass expansion, which we believe leads to results that approximate the full 2-loop $hgg$ and~$h\gamma\gamma$  form factors well. 

Assuming that~$\bar c_6$ is the only Wilson coefficient that receives a non-vanishing correction at the scale where new physics enters, we have analysed the sensitivity of present and future measurements of the signal strengths in  $gg \to h$ and $h \to \gamma \gamma$. In particular, we have demonstrated that the indirect constraints on $\bar c_6$ that follow from a combination of the~LHC~Run~I measurements of Higgs production in gluon-gluon-fusion and diphoton Higgs decays are more stringent than a direct extraction that uses the recent ATLAS upper limit on double-Higgs production. Our novel   95\%~CL  bound of $\bar c_6 \in [-12.7, 9.9]$ implies that  the trilinear Higgs coupling can at the moment still deviate from its SM value by a factor of approximately~11. 

We have furthermore investigated the prospects of  the  indirect constraints at future high-energy colliders. In the case of the HL-LHC with $3 \, {\rm ab}^{-1}$ of integrated luminosity, we have found that it should be possible to improve the present bound by a factor of more than~1.5, while for a FCC-ee an improvement by a factor of~2.5 seems feasible. The indirect tests proposed in our work could thus become sensitive to~$|\bar c_6| \simeq 5$, while  studies of double-Higgs production at the LHC may ultimately allow to set a 95\% CL bound of $\bar c_6 \in [-0.9,1.6] \cup [4.5,6.9]$. The sensitivity of our proposal is hence not sufficient to compete with the constraints of  $pp \to 2h$ for what concerns the solution  close to~$\bar c_6 = 0$, but it might allow to remove parts of the  flipped-sign solution centred around~$\bar c_6 \simeq 5.7$.

While the proposed indirect probes of $\lambda$ can clearly not replace the direct extraction of the trilinear Higgs coupling  at the LHC through double-Higgs production, we believe that they may turn out to be very valuable when included into a global analysis of Wilson coefficients, because compared to the direct measurement they constrain different linear combinations of effective operators in the SM EFT. An extension of our analysis of indirect  probes to other Higgs measurements, electroweak precision observables or quark flavour physics thus seems worthwhile and will be considered elsewhere. 

\acknowledgments 
We thank Florian Goertz for valuable discussions concerning double-Higgs production at the HL-LHC. The work of MG has been supported by the STFC consolidated grant ST/L000431/1. UH acknowledges the hospitality and support of the CERN theory division. He also would like to thank the KITP in Santa Barbara for hospitality and acknowledges that this research was supported in part by the National Science Foundation under Grant No.~NSF~PHY11-25915. UH is finally grateful to the MITP in Mainz for its hospitality and its partial support during the completion of this work.

\begin{appendix}

\section{Non-physical operators}
\label{app:EOM}

In addition to the gauge-invariant operators (\ref{eq:SILHoperator}), non-physical operators arise as counterterms in the renormalisation of higher loop one-particle-irreducible off-shell Green's functions with an insertion of the operator $O_6$. These non-physical operators can in general be divided into operators that vanish by the use of the equations of motion (EOM), non-physical counterterms  that can be written as a Becchi-Rouet-Stora-Tyutin (BRST) variation of other operators, i.e.~so-called BRST-exact operators, and evanescent operators that vanish algebraically in $d = 4$ dimensions.

For the calculation of the 2-loop anomalous dimensions that describes the mixing of~$O_6$ into $O_k$, it turns out that BRST-exact and evanescent operators do not play a role, and that only a single EOM-vanishing operator is necessary. This operator can be written as
\beq \label{eq:N1}
\begin{split}
N_1 = H^{\dagger}H \, \bigg [ & H^{\dagger} \big[ D_{\mu} D^{\mu} H \big ] + \big [ D_{\mu} \left ( D^{\mu} H \right) ^{\dagger} \big ] H \\[1mm] & - m_h^2  \left ( 1 - \frac{3 \bar c_6}{4} \right ) H^{\dagger} H   + 4 \lambda  \left ( 1 - \frac{3 \bar c_6}{2} \right )  \big ( H^{\dagger} H \big )^2  \\[1mm] &  +  \left (  Y_u \, \bar Q_L  u_R \hspace{0.25mm} \tilde H +  Y_d \,\bar Q_L d_R  \hspace{0.25mm} H +  Y_\ell \,\bar L_L   \ell_R  \hspace{0.25mm} H + {\rm h.c.} \right )  \bigg ] \,.
\end{split}
\eeq 
The terms $\bar c_6$ appear here because $\lambda$ denotes the combination $m_h^2/(2 v^2)$ of the Higgs mass~$m_h$ and its VEV $v$ $\big($see (\ref{eq:l34})$\big)$ and not the coefficient multiplying the quartic coupling~$(H^\dagger H)^2$ entering the SM Higgs potential. At the order we are working the terms proportional to $\bar c_6$ do not contribute to (\ref{eq:twoloopADM}). Note that the EOM-vanishing operator~$N_1$ arises as a counterterm independently of the IR regularisation adopted in the computation of the 2-loop anomalous dimensions of~$O_6$. However, if the regularisation respects the underlying symmetry, and all the diagrams are calculated on-shell, non-physical operators have vanishing matrix elements. In this case the operator given in (\ref{eq:N1}) would not contribute to  the mixing of physical operators. If~the gauge symmetry is broken this is no longer the case, as graphs with insertions of non-physical operators will generally project onto physical operators. Since our  IR regularisation implies massive boson propagators, non-physical counterterms play a crucial role at intermediate stages of our anomalous dimensions calculation.

\section{Feynman rules}
\label{app:Ophi}

\begin{figure}[!t]
\begin{center}
\includegraphics[height=0.175 \textwidth]{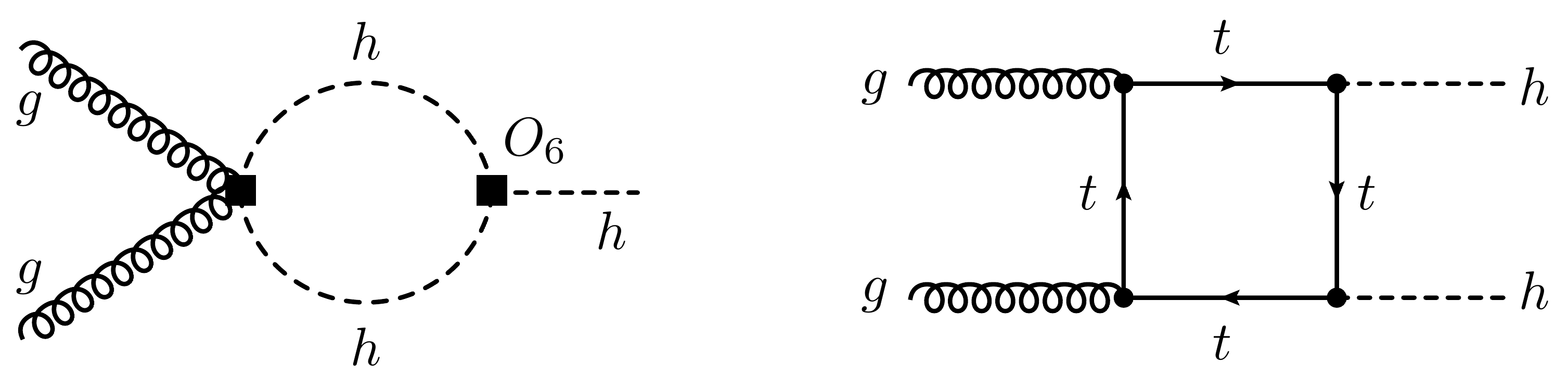}
\vspace{2mm}
\caption{\label{fig:gghrenormalisation} Left: UV divergent 1-loop diagram that involves the insertion of an effective~$gghh$ and~$h^3$ vertex. Right: An example of a 1-loop diagram that gives rise to an effective $gghh$ interaction after the top quark has been integrated out.}
\end{center}
\end{figure}

Inserting the explicit form of the SM Higgs doublet 
\beq \label{eq:phi}
H = \begin{pmatrix} \phi^+ \\[2mm] \frac{1}{\sqrt{2}} \left ( v + h + i \phi^0 \right ) \end{pmatrix} \,,
\eeq
into the expression for $O_6$ as given in (\ref{eq:SILHoperator}), one obtains the following interactions
\beq \label{eq:Ophi}
\begin{split}
O_6  & \supset - \left [  v h^3  + \frac{3}{2} \hspace{0.5mm} h^4 + \frac{3}{2} \hspace{0.5mm} h^2 \left ( (\phi^0)^2 + 2 \phi^+ \phi^- \right ) + \frac{3}{4v} \, h^5  \right. 
\\[2mm] & \left.  \phantom{xxxx}  + \frac{3}{2 v} \, h^3 \left ( (\phi^0)^2 + 2 \phi^+ \phi^- \right ) + \frac{3}{4 v} \, h \left ( (\phi^0)^2 + 2 \phi^+ \phi^- \right )^2  \right ] \lambda v^2  \,.
\end{split}
\eeq
Notice that $O_6$ does not contain a 3-point interaction of the form~$h \phi^+ \phi^-$, but 4-point interactions like $h^2 \phi^+ \phi^-$ and $h^2 \hspace{0.25mm} (\phi^0)^2$ as well as 5-point interactions such as $h^3 \phi^+ \phi^-$ and~$h \hspace{0.5mm} (\phi^0 )^2 \hspace{0.5mm}  \phi^+ \phi^-$. 

\section{Renormalisation procedure for $\bm{gg \to h}$ and $\bm{h \to \gamma \gamma}$}
\label{app:matching}

\begin{figure}[!t]
\begin{center}
\includegraphics[height=0.175 \textwidth]{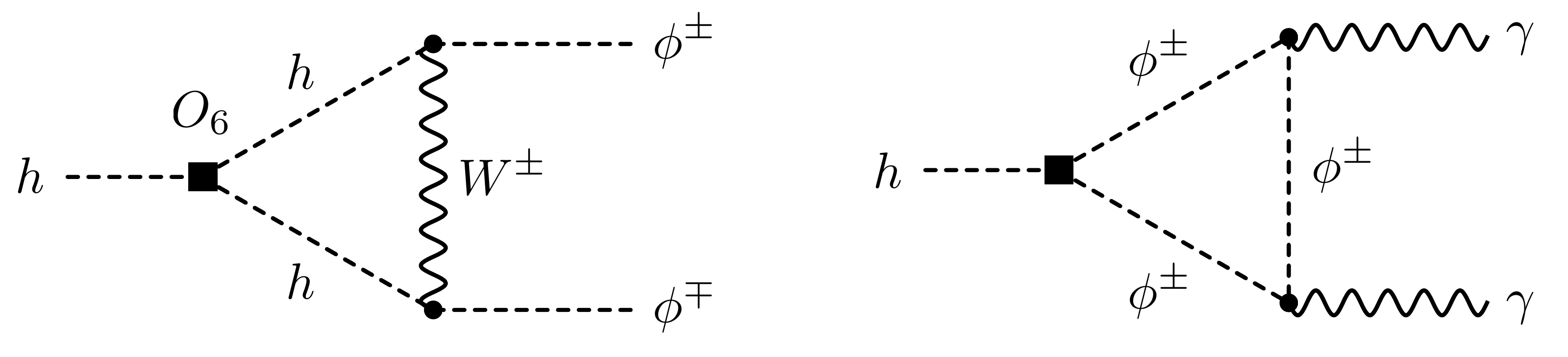}
\vspace{2mm}
\caption{\label{fig:haarenormalisation} Left: Diagram with an insertion of $O_6$ that leads to a UV divergent contribution to the $h\phi^+ \phi^-$ amplitude at the 1-loop level. Right: Feynman diagram that is needed to cancel the~UV poles of the 2-loop $h \to \gamma \gamma$   diagrams involving a divergent $h \phi^+ \phi^-$ subgraph. The black box indicates a counterterm insertion here.}
\end{center}
\end{figure}

In this appendix we briefly describe the renormalisation procedure that leads to our results~(\ref{eq:cg1}) and (\ref{eq:cA1}). We start with the ${\cal O} (\lambda)$ corrections to the $gg \to h$ amplitude. In the  limit of infinitely heavy top quark, we find the following unrenormalised 2-loop amplitude 
\beq \label{eq:Aggh1un}
{\cal A}^{(1)}_{\rm un}  ( gg \to h ) = \frac{\alpha_s}{\pi \hspace{0.25mm} v} \frac{\lambda \bar c_6}{(4 \pi)^2} \left [ -\frac{1}{4 \epsilon} - \frac{1}{2}  \ln \frac{\mu_w^2}{m_t^2} - \frac{25}{48} \right ] \,,
\eeq
where $\epsilon = (4-d)/2$. The renormalised 2-loop amplitude is obtained by adding to (\ref{eq:Aggh1un}) the counterterm  contribution  
\beq \label{eq:Aggh1ct}
{\cal A}^{(1)}_{\rm ct}  ( gg \to h )= \frac{\lambda \bar c_6}{(4 \pi)^2}   \frac{12 v}{\epsilon} \left \{ \frac{\alpha_s}{48 \pi \hspace{0.25mm} v^2} \left [ 1 + \epsilon \left ( \ln \frac{\mu_w^2}{m_t^2} + 2 \right) \right ] \right \}  \,.
\eeq
Here the first term proportional to $\bar c_6$ arises from the fact that for  $\tau_t \to \infty$ the operator~$O_6$ leads to a UV divergent $ggh$ matrix element through a 1-loop diagram involving an effective~$gghh$ coupling. The term in the curly brackets, on the other hand, represents the result for the 1-loop amplitude of $gg \to hh$ in the  infinite top-quark mass limit, including terms up to ${\cal O} (\epsilon)$. The corresponding Feynman graphs are depicted in Figure~\ref{fig:gghrenormalisation}. Notice that subtracting the counterterm contribution~(\ref{eq:Aggh1ct}) from the unrenormalised result~(\ref{eq:Aggh1un}), leads to a matching correction (\ref{eq:cg1}) that is independent of IR physics,~i.e.~our result for~$c_g^{(1)}$ does not depend on how light degrees of freedom are treated in the calculation. 

In the case of $h \to \gamma \gamma$, we instead find the following unrenormalised 2-loop amplitude 
\beq \label{eq:Ahaa1un}
{\cal A}^{(1)}_{\rm un}  ( h \to \gamma \gamma ) =   \frac{\alpha}{\pi \hspace{0.25mm} v} \frac{\lambda \bar c_6}{(4 \pi)^2} \left [ \frac{11}{4\epsilon} + \frac{11}{2}  \ln \frac{\mu_w^2}{m_W^2} +\frac{35}{8} \right ] \,,
\eeq
in the limit $\tau_W \to \infty$. The corresponding counterterm takes the form 
\beq \label{eq:Ahaa1ct}
\begin{split}
{\cal A}^{(1)}_{\rm ct}  ( h \to \gamma \gamma ) & =  \frac{\lambda \bar c_6}{(4 \pi)^2}   \frac{12 v}{\epsilon} \left \{ -\frac{7 \alpha}{32 \pi \hspace{0.25mm} v^2} \left [ 1 + \epsilon \left ( \ln \frac{\mu_w^2}{m_W^2} + \frac{44}{21} \right) \right ] \right \}  \\[2mm] 
& \phantom{xx}  + \frac{\alpha}{\pi} \frac{\lambda \bar c_6}{(4 \pi)^2} \, \frac{3 v}{2 \epsilon \hspace{0.25mm} s_w^2}  \left \{ -\frac{s_w^2}{12 \hspace{0.25mm} v^2} \left ( 1  + \epsilon  \ln \frac{\mu_w^2}{m_W^2}   \right ) \right \}  \,,
\end{split}
\eeq
where $s_w$ denotes the sine of the weak mixing angle. The counterterm contribution has been split into two parts. The first one represents (in full analogy to the case of $gg \to h$) the product of the UV divergent 1-loop matrix element involving $O_6$ and an effective $hh\gamma \gamma$ vertex times the gauge-boson contribution to the 1-loop $hh \to \gamma \gamma$ amplitude, evaluated for $m_W \gg m_h$. The second term subtracts the  UV divergences of the 2-loop diagrams that contain a divergent $h \phi^+ \phi^-$ subgraph. The relevant 1-loop diagrams needed for this subtraction are shown Figure~\ref{fig:haarenormalisation}. The graph on the left-hand side leads to the UV pole  in the second line of (\ref{eq:Ahaa1ct}), while the right diagram gives rise to the expression inside the curly bracket. 

\section{Unphysical Higgs sector}
\label{app:goldstones}

\begin{figure}[!t]
\begin{center}
\includegraphics[height=0.175 \textwidth]{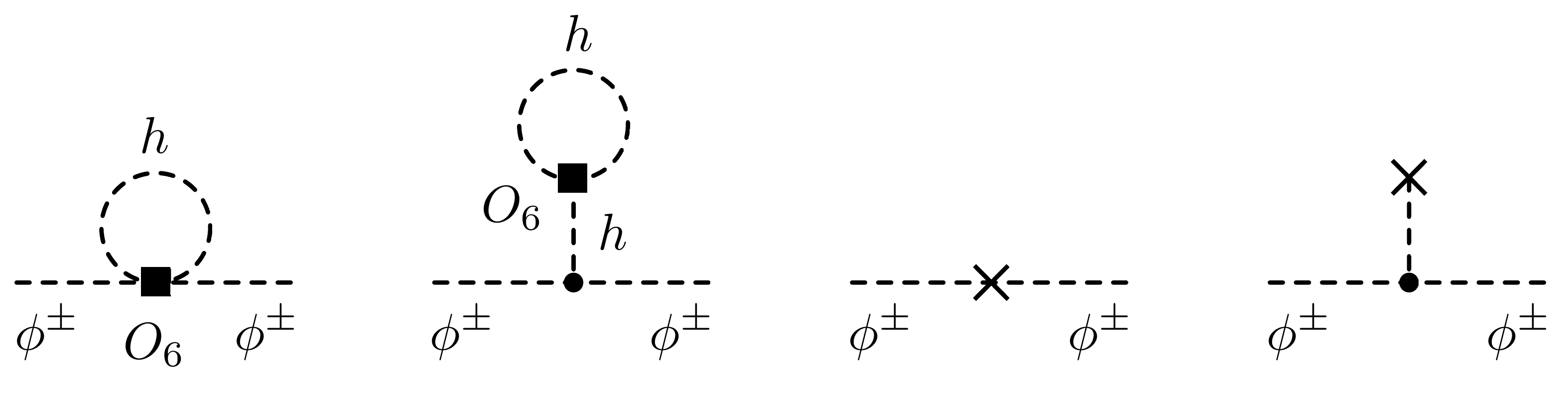}
\vspace{2mm}
\caption{\label{fig:goldstones} Feynman graphs with an insertion of $O_6$ (black box) that contribute to the selfenergy of the charged would-be Goldstone bosons $\phi^\pm$ at the 1-loop level. The corresponding counterterm~(black cross) diagrams are also shown.}
\end{center}
\end{figure}

To first order in $\lambda$ the charged would-be Goldstone boson propagator receives corrections from the Feynman diagrams shown in Figure~\ref{fig:goldstones}, and  similar graphs involving a virtual exchange of a neutral would-be Goldstone boson $\phi^0$. The calculation of the renormalised 1-loop Goldstone selfenergies requires a renormalisation condition that involves tadpoles. The most common ways to fix the tadpole contributions consists in requiring that the effective potential contains no term linear in the Higgs field~\cite{Denner:1991kt}  or in employing the $\overline{\rm MS}$ scheme~\cite{Jegerlehner:2001fb,Jegerlehner:2002em}. After a simple calculation it turns out that the ${\cal O} (\lambda)$ corrections to the Goldstone selfenergies cancel irrespectively of the precise treatment of the tadpoles. This is a result of gauge invariance.

The vanishing of the ${\cal O} (\lambda)$ corrections to the $\phi^\pm$ selfenergies implies that in the calculation of the coefficient $c_\gamma^{(1)}$ $\big($see (\ref{eq:cAexpansion})$\big)$ one does not have to consider 2-loop diagrams that involve a 1-loop correction to charged would-be Goldstone boson  propagators. One can furthermore show that 2-loop  ${\cal O} (\lambda)$ contributions that arise from the $h^3 \phi^+ \phi^-$ or $h \hspace{0.5mm} (\phi^0 )^2 \hspace{0.5mm}  \phi^+ \phi^-$ parts of $O_6$ $\big($see (\ref{eq:Ophi})$\big)$ are cancelled by 1-loop counterterm contributions, and that this cancellation is again independent of the precise treatment of the unphysical Higgs sector, as long as the procedure respects gauge invariance. 

\end{appendix}

\end{document}